\documentclass[prb,twocolumn,superscriptaddress,floatfix,aps,showpacs]{revtex4-1}
\usepackage[toc,page]{appendix}
\usepackage{graphicx,epsf,epsfig,amsmath}
\usepackage{alltt,dsfont,amsmath,amssymb,bm}
\usepackage{comment}

\usepackage{hyperref}
\usepackage{color}

\newcommand{\G}{{\cal{G}}}

\newcommand{\X}[2]{X_{{#1}}^{#2}}\label{fig:delta}

\newcommand{\ch}[1]{\hat{C}_{{#1} } }
\newcommand{\chdag}[1]{\hat{C}^\dagger_{{#1} } }

\newcommand{\si}{\sigma}

\newcommand{\tJ}{$t$-$J$\ }

\newcommand{\beq}{\begin{equation}}
\newcommand{\eeq}{\end{equation}}
\newcommand{\barray}{\begin{eqnarray}}
\newcommand{\earray}{\end{eqnarray}}
\newcommand{\disp}[1]{Eq.~(\ref{#1})}
\newcommand{\refdisp}[1]{Ref.~(\onlinecite{#1})}

\begin{document}
\title{Electronic spectral properties of the two-dimensional infinite-$U$ Hubbard model}
\author{Ehsan Khatami}
\affiliation{Physics Department, University of California, Santa Cruz, California 95064, USA}
\affiliation{Department of Physics, Georgetown University, Washington DC, 20057 USA}
\author{Daniel Hansen}
\affiliation{Physics Department, University of California, Santa Cruz, California 95064, USA}
\author{Edward Perepelitsky}
\affiliation{Physics Department, University of California, Santa Cruz, California 95064, USA}
\author{Marcos Rigol}
\affiliation{Department of Physics, The Pennsylvania State University, University Park,
Pennsylvania 16802, USA}
\author{B. Sriram Shastry}
\affiliation{Physics Department, University of California, Santa Cruz, California 95064, USA}


\begin{abstract}
A strong-coupling series expansion for the Green's function and the
extremely correlated Fermi liquid (ECFL) theory are used to calculate the
moments of the electronic spectral functions of the infinite-$U$ Hubbard model.
Results from these two complementary methods agree very well at both low
densities, where the ECFL solution is the most accurate, and at high to
intermediate temperatures, where the series converge. We find that a modified
first moment, which underestimates the contributions from the occupied states
and is accessible in the series through the time-dependent Green's function,
best describes the peak location of the spectral function in the strongly correlated
regime. This is examined by the ECFL results at low temperatures, where it is
shown that the spectral function is largely skewed towards the occupied
states.
\end{abstract}

\maketitle

\section{Introduction}
A long-standing theme in the dynamics of strongly interacting systems is the
reconstruction of  dynamics from the knowledge of the first few
moments.~\cite{h_mori_65} Its appeal lies in the relative ease with which these
moments can be computed, in contrast to computing the complete dynamical
correlation functions. The method of moments works well in cases where the
qualitative features of the correlation functions are somewhat understood by
other arguments, including conservation laws in the case of spin dynamics. In
the important problem of the strong-coupling Hubbard model, the moments are
dominated by the energy scale $U$,~\cite{w_nolting_72} the on-site repulsive
Coulomb interaction, and hence rendered useless.  In contrast, for the \tJ model
embodying extreme correlations, i.e., $U \to \infty$ at the very outset, a
better prospect exists. The moments are blind to the scale of $U$, since it does
not occur in the Hamiltonian, and therefore  one expects them to be meaningful
in determining the broad features of the dynamics. With this in mind, we study a
simple version of the \tJ model by focusing on $J=0$, which is identical to the
$U= \infty$ Hubbard model, thereby making more tools available for the analysis.
As we show in what follows, we have developed the capability to compute the
moments of the electron spectral function of this model by utilizing series
expansions.~\cite{domb,series} Experiments using angle-resolved photoemission
spectroscopy (ARPES)~\cite{gweon,johnson,zx,jc} directly measure this spectral 
function, providing an added impetus. 

An independent source of information about the electronic spectral function 
is the  recent analytical theory of extremely correlated Fermi liquids (ECFL). 
This theory has been developed in recent
publications,~\cite{ecfl,hansen-shastry} and several results of the model
pertaining to the detailed line shapes find close agreement with
experiment.~\cite{gweon} On the calculational front, the theory provides a
systematic methodology for computation, and the initial low order implementation
yields the single-electron spectral function for particle densities in the range
$0\leq n \lesssim 0.7$. The line shapes of this calculation for $n \geq 0.5$
display a characteristic skewed shape found in the experimental curves in ARPES,
as detailed in Ref.~\onlinecite{hansen-shastry}. The computed spectra are
available at any temperature (high or low), and the only limitation at present
is the inability to access the regime close to half filling with density 
greater than $n\sim0.75$. Given the inherent complexity of the newly developed
ECFL formalism, the possibility of an objective cross-check using series
expansions is a very attractive one, and here we provide the first  comparison.

We compute and compare the moments of the \tJ model with $J=0$ in two dimensions
by utilizing a series expansion~\cite{ehsan-method-paper} and the ECFL theory.
The two techniques are largely complementary. While they individually run into
difficulties in different regimes, namely, at low temperatures for the series
expansion and high densities for the ECFL, there is sufficient overlap in
densities and temperatures where {\em both methods} give reliable results. This 
provides  us with a unique opportunity to test the validity of the answers. For
ECFL, this provides a stringent test of the resulting moments by comparing with
the series expansion. For the series expansion, the availability of an
analytical theory and hence, of the entire spectrum, is of great advantage in
interpreting the distinctions between three  types of moments that can be
computed [see \disp{moments} below]. We find that especially at high densities,
the line shape of the spectral function is skewed towards occupied energies,
$\omega \leq 0$, therefore the spectral peak (SP) location (the maximum location 
in the energy distributed curves) is best estimated by the first moment of a modified 
function with dominant contribution from unoccupied states.

In the rest of this Rapid Communication, we first explain how the series
expansion and ECFL results are obtained (Sec.~\ref{sec:prelim}). In
Sec.~\ref{sec:results}, we compare the results from the two methods, and discuss
our findings. A summary follows in Sec.~\ref{sec:summary}.


\section{Preliminaries  }
\label{sec:prelim}

\subsection{Definitions of computed coefficients}

We  denote the imaginary-time Green's function for the $U = \infty$ Hubbard
model, or equivalently, the \tJ model with $J=0$, as  $\G(i, \tau_i; j, \tau_j)=
- \ \langle  T_\tau \hat{C}_{i \si}(\tau_i) \
\hat{C}^\dagger_{j \si} (\tau_j) \rangle$, where $T_\tau$ is the time-ordering
operator, and $\left<..\right>$ denotes the thermal 
expectation value. We thus study the limit of extreme correlations. The
operators are
Gutzwiller-projected Fermi objects and related to the Hubbard $X$ operators as
$\hat{C}_{i \si} \equiv \X{i}{0 \si }$, etc.  As usual,\cite{agd} this object is
a function of the time difference $\tau \equiv \tau_i- \tau_j$, and we will
study its spatial Fourier transform $\G(k,\tau)$. Our study begins with the
following expansions
\barray
\mathcal{G}(k,\tau>0)&=&(-1)\sum_{m=0}^{\infty}(-1)^m\frac{\tau^m}{m!}a_m(k),
\label{tpositive} \\
\mathcal{G}(k,\tau<0)&=&\sum_{m=0}^{\infty}(-1)^m\frac{\tau^m}{m!}b_m(k),
\label{tnegative}
\earray
where the coefficients $a_m$ are computed analytically as a series in the
hopping amplitude $t$. The series expansion can be carried out to the fourth
order by hand,~\cite{edward-method-paper} and pushed to the eighth order by a
highly efficient computer program~\cite{ehsan-method-paper} based on Metzner's
linked-cluster formalism.~\cite{Metzner} This order is the limit achievable by
currently available supercomputers. Using antiperiodic boundary conditions,
$\G(\tau - \beta) = -\G(\tau)$, we obtain \disp{tnegative} from
\disp{tpositive}. Here $\beta=1/(k_B T)$ is the inverse temperature (we set $t=1$
as the unit of energy, and $k_B=1$). Therefore, the main calculation
focuses on \disp{tpositive}. Its Fourier series in Matsubara frequencies,
$\omega_n=(2n+1)\pi/\beta$, is obtained from $\G(k,i \omega_n) = \int_0^\beta
e^{ i \omega_n \tau}\,\G(k,\tau)\, d \tau $. The spectral function at momentum
$k$ and for the real frequency $\nu$ is denoted by $\rho_\G(k,\nu)$ and
determines the Green's function through the relation $
\G(k,i\omega_n)=\int_{-\infty}^{+\infty}\frac{\rho_{\G}(k,\nu)}{i\omega_n-\nu}
d\nu $. At high frequencies $\omega_n$, we have an expansion
\begin{equation}
\G(k,i\omega_n)=\sum_{m=0}^{\infty} \frac{c_m(k)}{(i\omega_n)^{m+1}},\nonumber
\end{equation}
involving the ``symmetric'' coefficient, $c_m(k)$ (see below).
The time domain Green's function is also given
in terms of the spectral function
by the important  representation
\begin{equation}
\mathcal{G}(k,\tau)=\int_{-\infty}^{+\infty}d\nu \rho_{\G}(k,\nu)e^{-\nu
\tau}\left[\Theta(-\tau)f(\nu)-\Theta(\tau)\bar{f}(\nu)\right],
\end{equation}
where
\begin{equation}
f(\nu)=\frac{1}{1+e^{\beta \nu}}\ \ \ \textrm{and}\ \ 
\bar{f}(\nu)=\frac{1}{1+e^{-\beta \nu}}.
\end{equation}
The three sets of coefficients $\alpha_m$ (i.e., $a_m$, $b_m$, and $c_m$) are
easily seen to originate from the spectral function convoluted by a different
filter function $\chi(\nu)$ [respectively $\bar{f}(\nu), f(\nu), 1$] as
\beq
\alpha_m(k) = \int_{-\infty}^\infty \nu^m \chi(\nu) \rho_{\G}(k,\nu) \ d\nu.
\label{alpham}
\eeq
Using this and the identity $f+\bar{f}=1$, we see that the symmetric
coefficients satisfy the important relation
\beq c_m(k)=a_m(k)+b_m(k) \label{addition}. \eeq

\subsection{Definition of moments}

Equation \eqref{alpham} gives the power integrals of the effective
spectral function $\chi(\nu) \rho_{\G}(\nu)$, and naturally leads to
three  sets of moments at each $k$,  $\varepsilon^{\chi}_{\
m}(k)=\alpha_m(k)/\alpha_0(k)$. Thus, the moments can be obtained from the
coefficients $a_m,b_m,c_m$, and contain complementary information as
we discuss below.  We assign them suggestive names 
\beq
\varepsilon_m^>(k)=\frac{a_m(k)}{a_0(k)}, \;\;\;
\varepsilon_m^<(k)=\frac{b_m(k)}{b_0(k)},
\;\;\;\varepsilon_m^0(k)=\frac{c_m(k)}{c_0(k)}, \label{moments}
\eeq
the {\em greater}, {\em lesser}, and {\em symmetric} moments,
respectively.~\cite{farid} The superscripts in the notations $\varepsilon^>$ and
$\varepsilon^<$ signify that the contribution to these energy moments  comes
predominantly from the weight of the spectral function that lies {\em above} or
{\em below} the chemical potential, and hence the unoccupied or occupied states.
The coefficients at $m=0$ have special meanings: By computing the
anticommutator of $\hat{C}$ and $\hat{C}^\dagger$, and taking its average we
find $c_0(k) \equiv c_0=1-\frac{n}{2}$ in this model. The coefficient $b_0(k)$
is also the
momentum distribution function,
\beq
m_\si(k)= \langle \chdag{k \si} \ch{k \si}  \rangle = b_0(k).
\eeq
Using \disp{addition}, we find $a_0(k)= 1- \frac{n}{2} - m(k)$.

In this work, we study  only the first moments, i.e., $m=1$. We argue below
that these  give an estimate of the quasiparticle  spectrum  for a given $k$. It
is particularly useful to study all three moments separately since they exhibit
different behavior, and the comparison with the spectra of ECFL gives a clearer
understanding of their differences, as we discuss below.

\subsection{Summary of relevant ECFL results}

In Ref.~\onlinecite{hansen-shastry}, the formalism of ECFL for general $J$ 
is implemented to
second order in the variable $\lambda$, which is closely related to the density.
A self-consistent argument indicates that the calculation in
\refdisp{hansen-shastry} is 
 valid for densities $n \lesssim 0.7$. It has no limitation on the
temperature or system size, since it is essentially an analytical theory
--resembling the skeleton graph expansion theories of standard models in
structure. We note that the ECFL assumes a specific type of Fermi liquid with 
strong asymmetric corrections,~\cite{ecfl} and the reasonable similarity to the series 
data, as we will see in Sec.~\ref{sec:results}, suggests that this conclusion is 
fairly safe, at least for high enough temperatures. At low temperatures, there 
could be other instabilities that are hard to capture with the series analysis, 
and the present versions of the ECFL.

The full spectral function $\rho_{\G}(k,\nu)$ is computed and its
moments (for the case of $J=0$) are readily available for comparison with those 
from the series expansion. Also available in this work is the location of the SPs
$\varepsilon^{\textrm{SP}}(k)$, when they exist, the momentum distribution function, 
etc. It is therefore possible to compute various dispersion curves, relating the
different characteristic energies (i.e., moments) to wave vectors, and to
compare them with the true SP dispersion. The benchmarking of these moments
provides us with valuable insight for interpreting the series data, where the
SPs are not available, but the moments are.


\vspace{-0.5cm}
\section{Results}
\label{sec:results}

\begin{figure}[!t]
\includegraphics[width=0.48\textwidth]{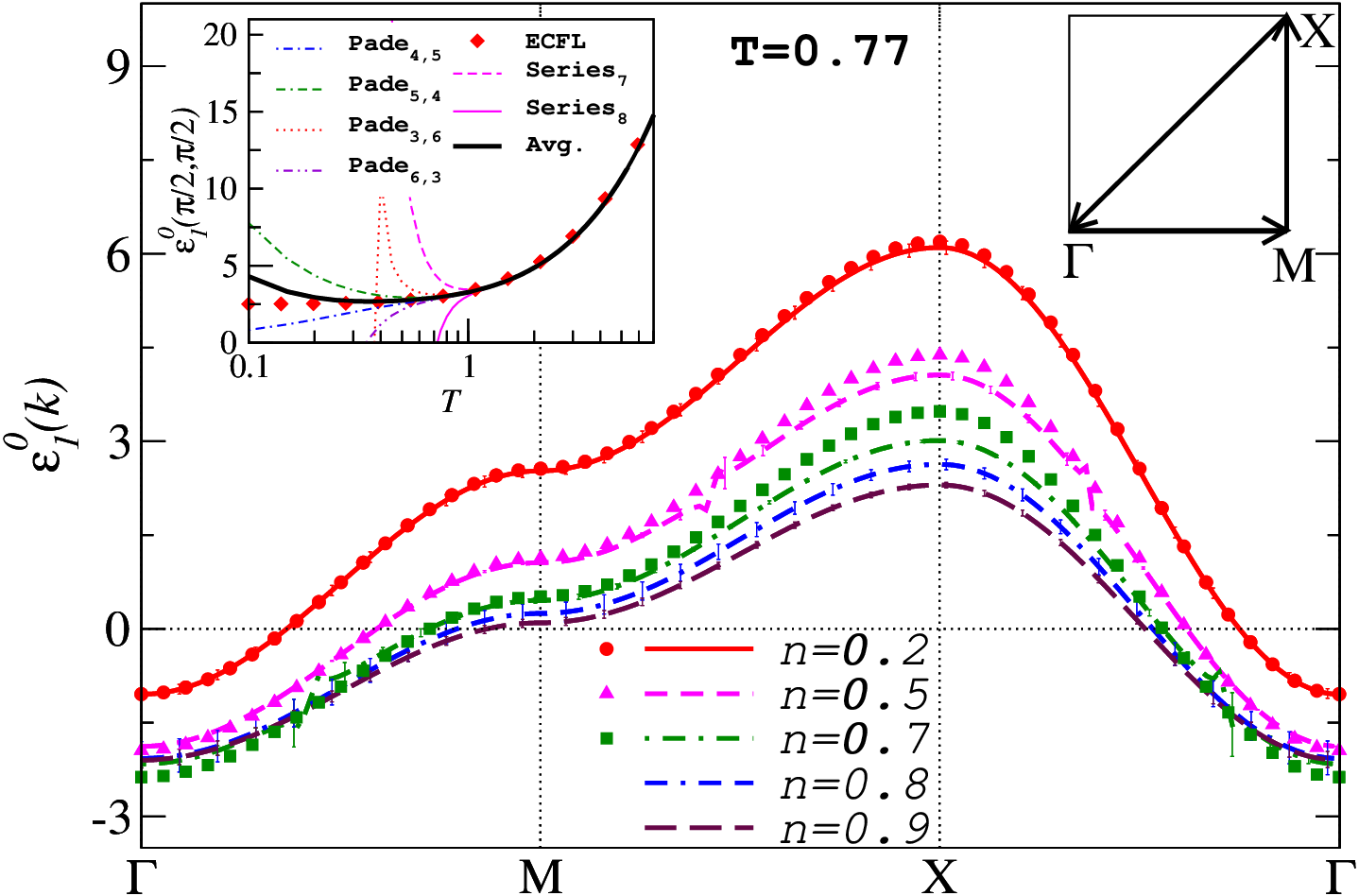}
\caption{(Color online) The first symmetric moment $\varepsilon_1^0(k)$ at $T=0.77$
vs momentum around the irreducible wedge of the Brillouin zone (the path is
shown in the right inset). Lines are results from the series and symbols for
$n\leq0.7$ are from ECFL calculations. Left inset: $\varepsilon_1^0(k)$ for
$n=0.2$ at 
$k=(\pi/2,\pi/2)$ from the ECFL (diamonds), up to orders seven and eight of the
series (labeled Series$_7$ and Series$_8$), and up to the eighth order after
various Pad\'{e} approximations, vs temperature on a logarithmic scale. The numbers
in the subscripts of ``Pad\'{e}'' labels represent the order of the polynomial in
the numerator and in the denominator of the Pad\'{e} ratio, respectively. ``Avg.''
denotes the average between Pad\'{e}$_{\{4,5\}}$ and Pad\'{e}$_{\{5,4\}}$. In the main
panel, the results for the series are either the average between
Pad\'{e}$_{\{4,5\}}$ and Pad\'{e}$_{\{5,4\}}$ or Pad\'{e}$_{\{5,5\}}$ and 
Pad\'{e}$_{\{5,4\}}$,~\cite{pade} with the ``error bars'' defined as the differences
between the two.~\cite{errorbars}}
\label{fig:First}
\end{figure}

In Fig.~\ref{fig:First}, we plot the symmetric first moment
$\varepsilon_1^0(k)$ as a function of momentum at $T=0.77$ for five
different densities $n=0.2$, 0.5, 0.7, 0.8, and 0.9. We find excellent
agreement between the results from the series and the ECFL for $n=0.2$ for all
the momenta around the irreducible wedge of the Brillouin zone. At higher
densities up to $n=0.7$ (beyond which the ECFL results are not quoted), the
agreement is still very good, except around the zone corner, where the
disagreement grows as the density increases.~\cite{shift} The results for the
series are obtained from Pad\'{e} approximations as the bare results show divergent
behavior at $T<1$. The number of terms in the series is large enough to justify
the utilization of Pad\'{e} approximations in order to extend the convergence to lower
temperatures. A comparison of several of these approximations with the ECFL
results for a (low) density of $n=0.2$ is shown in the inset of
Fig.~\ref{fig:First}. In that case, we see that the agreement between the two
methods extends to temperatures as low as $T=0.3$ using Pad\'{e} approximations.

\begin{figure}[!t]
\includegraphics[width=0.48\textwidth]{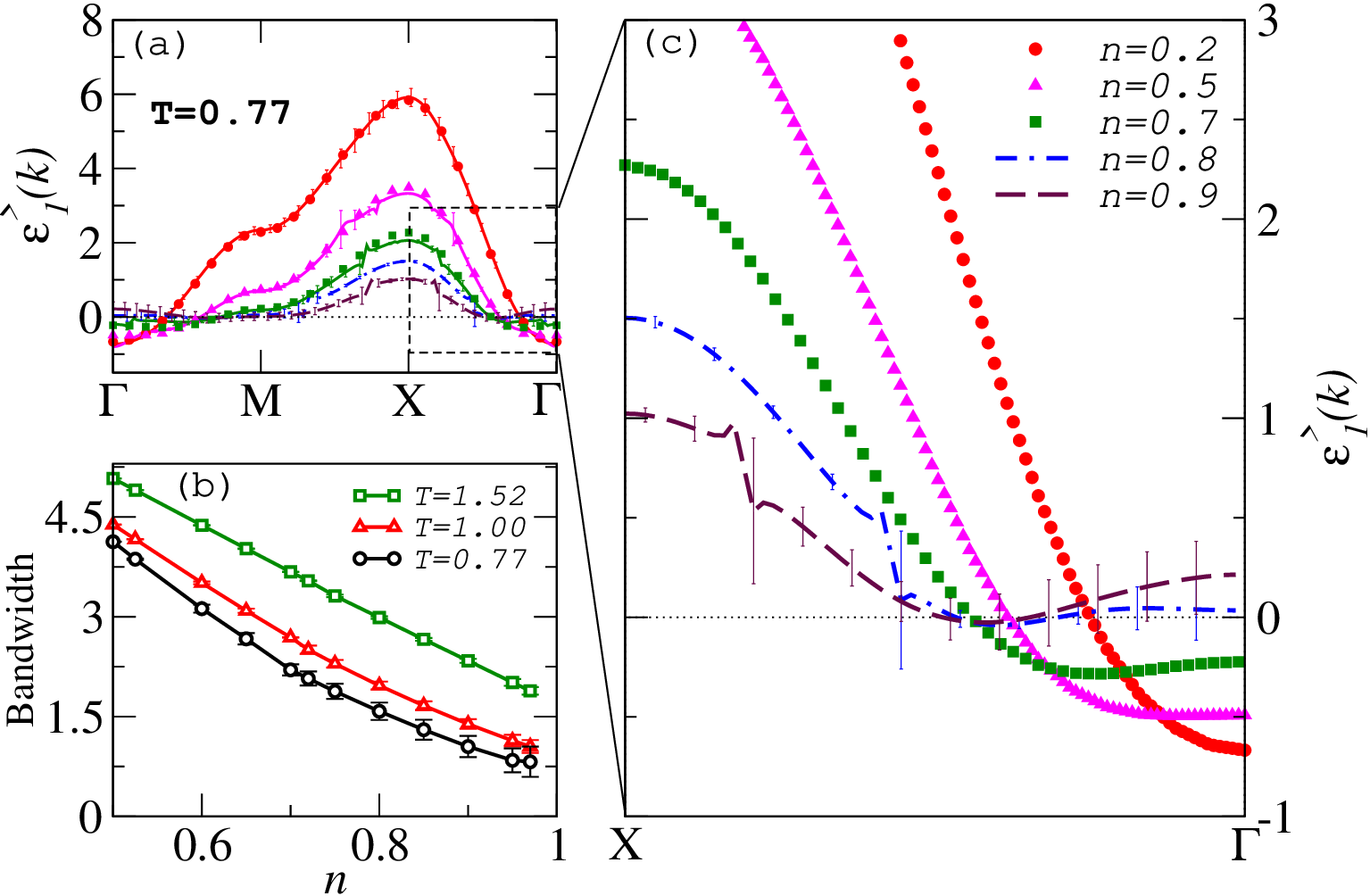}
\caption{(Color online) (a) The first {\em greater} moment $\varepsilon_1^>(k)$
at $T=0.77$ vs momentum for the same path around the irreducible wedge of the
Brillouin zone as in Fig.~\ref{fig:First}. Lines and symbols are also the same
as in Fig.~\ref{fig:First}. (b) The bandwidth of $\varepsilon_1^>(k)$, defined
as the difference between its maximum and minimum values at momenta shown in
panel (a), vs density for $T=1.52, 1.00$, and $0.77$. Panel (c) zooms in the
results in panel (a) for $k$ along the nodal direction. The two methods
more or less agree with each other, within the error bars, in this window for
$n\leq0.7$, and therefore, we show only the ECFL results for the latter cases.}
\label{fig:FirstGreater}
\end{figure}

The greater moment $\varepsilon_1^>(k)$ is plotted in Fig.~\ref{fig:FirstGreater}(a) 
at the same temperature and densities as in Fig.~\ref{fig:First}. 
For $\varepsilon_1^>(k)$, the overall agreement between the series expansions and the 
ECFL results for all $n\leq0.7$ is better than for $\varepsilon_1^0(k)$, especially 
around the $X$ point. We also note that $\varepsilon_1^>(k)$ exhibits a more intriguing 
behavior than $\varepsilon_1^0(k)$. One of the prominent features of the former, seen 
in Fig.~\ref{fig:FirstGreater}(a), is the significant narrowing of the band
by increasing the density. In Fig.~\ref{fig:FirstGreater}(b), we plot the
bandwidth [i.e., $\max(\varepsilon^>_{1})-\min(\varepsilon^>_{1})$]
from the series as a function of density at $T=1.52$, 1.00, and 0.77. It
appears that the bandwidth deviates from a linear dependence on $n$ by
decreasing the temperature, and saturates for $n\to 1$ at a nonzero value that
decreases towards zero with decreasing $T$. Close to $n=1$ at $T=0.77$, we find
a weaker agreement between different Pad\'{e} approximations, leading to larger error
bars. The version of ECFL in \refdisp{hansen-shastry} cannot be used to study
this effect as the high-density region $n \sim 1$ is beyond its regime of
validity.

\begin{figure}[!t]
\includegraphics[width=0.48\textwidth]{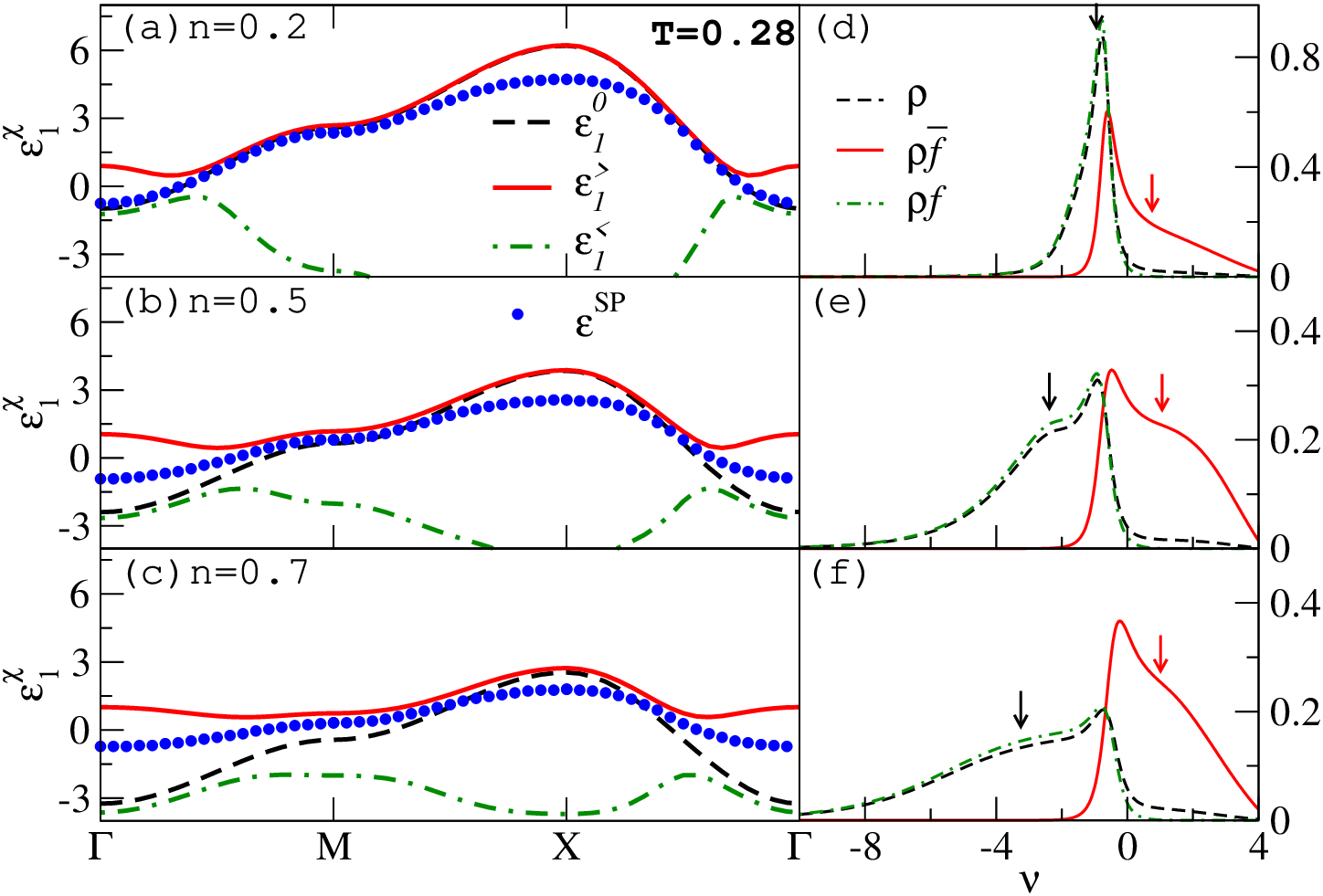}
\caption{(Color online) Comparison of the SP location $\varepsilon^{\textrm{SP}}(k)$ (symbols)
and the three moments from ECFL at $T=0.28$ and for (a) $n=0.2$, (b) $n=0.5$, and (c)
$n=0.7$. Right panels show the corresponding spectral functions and their
products to $\bar{f}(\omega)$ and
$f(\omega)$ at $\Gamma$ for the same densities shown in the left panels. 
Dark (Light) arrows show the values of $\varepsilon_1^0$ ($\varepsilon_1^>$). 
At low densities, the SP location is estimated well by the first symmetric
moment. At higher density, the spectral function is skewed and the greater
moment, which is calculated for the spectral function after most of its weight
in the negative frequency region is cut off, provides a better estimate.}
\label{fig:qppeak}
\end{figure}

Another interesting feature of $\varepsilon_1^>(k)$ [Fig.~\ref{fig:FirstGreater}(a)]
is the change in sign of its slope near the $\Gamma$ point as the density increases 
towards unity. To better study this feature, in Fig.~\ref{fig:FirstGreater}(c), 
we report only the results along the nodal direction. We find that for
$n\gtrsim0.7$, the greater moment initially decreases as the momentum increases from
zero, leading to a negative curvature, or  effective mass, at the $\Gamma$
point. This feature becomes more pronounced as we increase the
density, or decrease the temperature (see Fig.~\ref{fig:qppeak}). These results
hint at  a possible reconstruction of the Fermi surface, i.e.,  the negative mass
persisting and extending in $k$ space so as to reach the Fermi momentum.  The
appearance of such a hole pocket in the (hole) underdoped regime, could be of 
interest in ARPES and quantum oscillation studies. However, establishing this 
firmly requires higher order terms in the series, and is therefore difficult.

So far, we have seen that for intermediate temperatures and at relatively small
densities, the ECFL agrees extremely well with the results of the series
expansion. But, unlike the series expansion,
ECFL is not limited to high temperatures at those densities and can be used to
study the moments, and more importantly, the real-frequency spectral functions,
at much lower temperatures. Therefore, we focus on the ECFL results at $n=0.2$,
0.5, and 0.7, and at a reduced temperature of $T=0.28$, a temperature at which the
series do not converge. In Figs.~\ref{fig:qppeak}(a)-\ref{fig:qppeak}(c),
we plot $\varepsilon_1^0(k)$, $\varepsilon_1^>(k)$, and $\varepsilon_1^<(k)$ from
the ECFL, along with $\varepsilon^{\textrm{SP}}(k)$, obtained from the spectral
functions, at different momenta. We find that in the physically interesting
region of low temperatures and high densities, where correlation effects are
strongest, the location of the SP is generally better estimated by the
greater moment than by the symmetric, or the lesser one [see
Fig.~\ref{fig:qppeak}(c)].

The spectral functions shown in Figs.~\ref{fig:qppeak}(d)-\ref{fig:qppeak}(f)
help us understand why this is the case. There, we plot the spectral
functions $\rho_{\G}(k,\omega)$, $\rho_{\G}(k,\omega)\bar{f}(\omega)$, and
$\rho_{\G}(k,\omega)f(\omega)$, corresponding to the three moments at
$k=(0,0)$, where the differences between the moments are the most pronounced, vs
frequency. At $n=0.2$, there exists a relatively sharp quasiparticle peak in $\rho_{\G}$
whose location matches the first symmetric moment (marked by a dark arrow) very
well. $\varepsilon_1^>(k)$, on the other hand, falls slightly to the right of
the quasiparticle peak (marked by a light-colored arrow) as most of the spectral weight in
negative frequencies is cut off after multiplying $\rho_{\G}$ by
$\bar{f}(\omega)$ [see Eq.~\eqref{alpham}]. Also, since there is very little
spectral weight in the positive frequency side, $\varepsilon_1^<(k)$ is very
close in value to $\varepsilon_1^0(k)$. As the density is increased to $n=0.5$,
the spectral function is skewed as a result of correlations. In this case, at
small $k$, there is much more spectral weight on the left of the SP than on
the right, causing the symmetric moment to be smaller than
$\varepsilon^{\textrm{SP}}(k)$. This feature becomes more significant at a higher density
of $n=0.7$, where almost all of the spectral weight is in the negative frequency
side. As a result, multiplying  $\rho_{\G}$ by $\bar{f}(\omega)$ helps in
neglecting the excess weight on the left side of the SP. Hence,
$\varepsilon_1^>(k)$, which is readily available from the series at even higher
densities, may be used as an indicator of $\varepsilon^{\textrm{SP}}(k)$ using this
insight from the ECFL spectra.


\section{Summary}
\label{sec:summary}

We employ two complementary methods, namely, a strong-coupling series expansion
and the ECFL, to calculate the moments of the spectral functions for the
infinite-$U$ Hubbard model.  Unveiling the basic physics of the model is 
benefited by the complementarity of those approaches. Furthermore, the series 
expansion results provide the first independent check of the ECFL theory, which
has been self-consistently established. 
At intermediate temperatures and low densities, where the
results from both methods are available, we find very good agreement between
the two. Unlike ECFL, the series is not limited to small densities and, by
increasing the density in the series to near half filling, we find interesting
features in the dispersion of the moment with dominant contributions from
unoccupied states (the greater moment). These include a significant narrowing of
its band as well as hints of Fermi-surface reconstruction. Unlike the series,
the ECFL is not limited to high temperatures and, by exploring the ECFL results
at lower temperatures, we find that the greater moment better describes the
location of the SP as the density increases. This is understood based on
the skewing of the spectral functions in the negative frequency region in the
strongly-correlated regime.


\section{Acknowledgments}
This work was supported by DOE under Grant No.~FG02-06ER46319 (B.S.S., D.H.,
and E.P.), and by NSF under Grant No.~OCI-0904597 (E.K. and M.R.).


\begin{thebibliography}{99}


\bibitem{h_mori_65}
H. Mori, Prog. Theor. Phys. {\bf 33}, 423 (1965); {\bf 34}, 399
(1965); M. Dupuis, {\it ibid.} {\bf 37}, 502 (1967); K. Tomita and H.
Mashiyama, {\it ibid.} {\bf 51}, 1312 (1974).



\bibitem{w_nolting_72} 
W. Nolting, Z. Phys. {\bf 255}, 25 (1972);
J. J. Deisz, D. W. Hess, and J. W. Serene, Phys. Rev. B {\bf 66}, 014539 (2002).

\bibitem{domb} 
{\em Phase Transitions and Critical Phenomena}, edited by C. Domb and M. S.
Green (Academic, London, 1974), Vol. 3.

\bibitem{series}
W. Brauneck, Z. Phys. B {\bf 28}, 291 (1977); K. Kubo, Prog. Theor.
Phys. {\bf 64}, 758 (1980); R. R. P. Singh and R. L. Glenister, Phys. Rev. B
{\bf 46}, 14313 (1992); {\bf 46}, 11871 (1992); W. O. Putikka, M. U.
Luchini, and T. M. Rice, Phys. Rev. Lett. {\bf 68}, 538 (1992); W. O.
Putikka, M. U. Luchini, and R. R. P. Singh, {\it ibid.} {\bf 81}, 2966
(1998).


\bibitem{gweon} G.-H. Gweon, B. S. Shastry, and G. D. Gu, Phys. Rev. Letts {\bf
107}, 056404 (2011); K. Matsuyama, G. H. Gweon, arXiv:1212.0299 (unpublished).

\bibitem{johnson} T. Valla,  A. V. Fedorov, P. D. Johnson, B. O. Wells, S. L.
Hulbert, Q. Li, G. D. Gu, N. Koshizuka, Science {\bf 285}, 2110 (1999).

\bibitem{zx} C. G. Olson et al., Phys. Rev. B {\bf 42}, 381 (1990), T. Yoshida
et al., J. Phys. Condens. Matter {\bf 19}, 125209 (2007).

\bibitem{jc}  A. Kaminski et al., Phys. Rev. B {\bf 69}, 212509 (2004)

\bibitem{ecfl} B. S. Shastry, Phys. Rev. Lett. {\bf 107}, 056403 (2011); 
{\bf 108}, 029702 (2012); Phys. Rev. B {\bf 84}, 165112
(2011); {\bf 87}, 125124 (2013); Phys. Rev. Lett. {\bf 109}, 067004 (2012); 
Phys. Rev. B {\bf 81}, 045121 (2010).

\bibitem{hansen-shastry} D. Hansen and B. S. Shastry, arXiv:1211.0594
(unpublished).

\bibitem{ehsan-method-paper} E. Khatami, E. Perepelitsky, B. S. Shastry, and M.
Rigol (unpublished).


\bibitem{agd} A. A. Abrikosov, L.  Gorkov and I. Dzyaloshinski, {\em Methods of
Quantum Field Theory in Statistical Physics},  (Prentice-Hall, Englewood
Cliffs, NJ, 1963).


\bibitem{edward-method-paper} E. Perepelitsky and B. S. Shastry (unpublished).


\bibitem{Metzner} W. Metzner, Phys. Rev. B {\bf 43}, 8549 (1991).


\bibitem{farid} 
B. Farid, Philos. Mag. {\bf 84}, No. 9, 909 (2004).


\bibitem{pade} Note that the zeroth order terms in the expansion for
$\varepsilon_1^0(k)$ is proportional to $\beta^{-1}$. Hence, Pad\'{e}$_{\{l,m\}}$
with $l+m=9$ will match the series order by order exactly up to the eighth
order. Nevertheless, we find that Pad\'{e}$_{\{5,5\}}$, for which one assumes that
the coefficient of the ninth order term in the series is zero, often results in
less spurious features than with Pad\'{e}$_{\{4,5\}}$, and therefore is used instead
of the latter for $n=0.2$ and $0.9$. On the other hand, since
$\varepsilon_1^>(k)$ is itself a ratio of two polynomials, either of the above
two Pad\'{e} approximants is equally valid. In this case
(Fig.~\ref{fig:FirstGreater}), we use the average of Pad\'{e}$_{\{5,4\}}$ and
Pad\'{e}$_{\{5,5\}}$ for $n=0.8$ and Pad\'{e}$_{\{5,4\}}$ and Pad\'{e}$_{\{4,5\}}$ for 
the rest.

\bibitem{errorbars} There is no error per se in the calculation of the
coefficients of terms in the series. The so-called error bars are merely a
measure of the convergence limit for the Pad\'{e} approximations at low
temperatures, where the bare results show divergent behavior. They do not
represent statistical or particular systematic errors.

\bibitem{shift} We may take the curves of  $\varepsilon_1^0(k)$, or more 
accurately, $\varepsilon_1^>(k)$ as estimates of the SP dispersion 
$\varepsilon^{\textrm{SP}}(k)$, after shifting them by a constant chosen to pass them 
through zero energy at the Fermi momentum (as in Figs.~\ref{fig:First} and 
\ref{fig:FirstGreater}). The magnitudes of the shift constants 
are on the scale seen in Figs.~\ref{fig:qppeak}(d)-\ref{fig:qppeak}(f) as the 
separation between the peak locating the  $\varepsilon^{\textrm{SP}}(k)$ and the arrows 
locating the moments.



\end{thebibliography}
\end{document}